\documentclass[aps,prd,twocolumn,nofootinbib]{revtex4-1}
\usepackage{epsfig}
\usepackage[colorlinks,linkcolor=blue,anchorcolor=blue,citecolor=blue,urlcolor=blue,breaklinks=true]{hyperref}
\usepackage{graphics}
\usepackage{color}
\usepackage{slashed}
\usepackage{makecell}
\usepackage{subfigure}
\usepackage{changes}
\usepackage{cancel}
\usepackage{amsthm,amsmath,amssymb}
\usepackage{mathrsfs}
\usepackage{dsfont}

\begin{document}
\author{Shu-Sheng Xu$^{1}$}~\email[]{Email: xuss@njupt.edu.cn}
\address{$^{1}$ School of Science, Nanjing University of Posts and Telecommunications, Nanjing 210023, China}

\title{QCD equation of state and the structure of neutron stars in NJL model}
\begin{abstract}
The chiral and pion superfluidity phase transitions are studied within NJL model. The model parameters are fitted by pion mass and decay constant together with recent isospin density from lattice QCD. The dense and cold QCD matter suffered the first order chiral phase transition during the quark chemical potential increase. Based on the studies of phase structures, the equation of states~(EoS) of dense QCD matter are calculated with the conditions of $\beta$ equilibrium and electric charge neutrality. Take the EoS as input, the mass-radius relation of neutron stars are predicted, which is consistent with recent observations of heaviest mass and radius of $1.4M_\odot$.

\bigskip


\bigskip

\end{abstract}
\maketitle

\section{Introduction}\label{intro}
The structures of compact stars are one of the most essential topics in astrophysics. They are extremely dense objects, forming from the collapse of massive stars, which possess smaller radius compared to ordinary stars. Light neutron stars may converted into quark stars further, and heavy neutron stars finally collapse into black holes. This topic is also attracted a lot interest from particle physicists, because of the compact stars are high density quantum chromodynamics~(QCD) systems with low temperature. It is an ideal place to study the phase structure of QCD.

The QCD equation of state~(EoS) at finite density and low temperature is crucial to the structures of compact stars~\cite{Ozel2006,Alford2007}. However, the phase structures of QCD at finite baryon chemical potential~($\mu_B$) cannot obtained from the first principle. Lattice QCD~(LQCD) is the lattice regularized of the QCD, which have been employed to study finite temperature~($T$) and finite isospin chemical potential~($\mu_I$) with $\mu_B=0$~\cite{Borsanyi2010,PhysRevLett.110.172001}, 
but at finite baryon chemical potential some extra difficulties emerge, namely due to the notorious ``sign problem''.
 In such a situation, people have to resort various effective models to study the phase structures of QCD at finite baryon chemical potential, such as chiral perturbation theory~\cite{PhysRevD.88.031502, AoP.158.142--210, NPB.250.465--516, PhysRevC.80.034909}, quasi-particle model~\cite{PhysRevD.82.014023,PhysRevD.84.094004,Szab2003,PhysRevC.79.055207}, Dyson-Schwinger equations~(DSEs)~\cite{PPNP.45.S1,GRUTER2004343,Maas:2005hs,PRL.106.172301,PhysRevD.88.014007,FISCHER20131036,PhysRevD.90.034022,PAWLOWSKI2014113, JHEP.07.014, PhysRevD.91.056003,Xu:2019ccc,FISCHER20191,PhysRevD.100.074011,PhysRevD.102.034027} and Nambu-Jona-Lasinio~(NJL) models~\cite{PPNP.27.195, RMP.64.649, EPJC.73.2612, EPJC.74.2782}. Some of them have been used to study neutron stars~\cite{PhysRevD.83.025012,PhysRevD.86.114028,PhysRevD.91.034018,PhysRevD.92.054012,PhysRevD.95.056018,PhysRevD.97.103013,PhysRevD.101.063023,PhysRevD.101.103021}.

For cold and dense QCD systems, nucleons appear at $\mu_B=\mu_B^0=m_N$, where $m_N$ is the mass of a nucleon. During the increase of $\mu_B$, the nuclear matter deconfined into quark matter, which is so-called deconfinement phase transition~\cite{stephanov2004qcd}. The dynamical chiral symmetry breaking may plays crucial role in deconfinement phase transitions~\cite{PhysRevLett.50.393,AHARONY20071420,Maris:2002mt}, which location is close to the location of chiral phase transition. On this understanding, the proton and neutron exist in the QCD matter when  $\mu_B$ between $\mu_B^0$ and the critical baryon chemical potential of deconfinement, $\mu_B^c$, and the quark matter appears when $\mu_B>\mu_B^c$.

Neutron stars are unique laboratories to test the effective models of dense QCD. For a specific central pressure, one can obtain a mass and radius by solving Tolman-Oppenheimer-Volkoff~(TOV) equation, which depends on EoS of nuclear matter. There are lots of informative pulsars which are helpful to rule out unrealistic EoSs. For example, PSR J0348+0432~(with mass $2.01\pm 0.04~M_\odot$)~\cite{Antoniadis1233232}, PSR J1614-2230~(with mass $1.97\pm 0.04~M_\odot$)~\cite{Arzoumanian_2018}. More recently, a mass of $2.14^{+0.10}_{-0.09}M_\odot$ (with 68.3\% credibility interval) for pulsar J0740+6620 was reported~\cite{cromartie2020relativistic}, which may replace the previously reported heaviest PSR J0348+0432. Besides these heavy neutron stars, PREX collaboration report the upper and lower limit radius of a $1.4M_\odot$ neutron star, that is $13.25~\mathrm{km}\lesssim R^{1.4M_\odot}\lesssim 14.26~\mathrm{km}$~\cite{PhysRevLett.126.172503}.

Effective model studies usually fit parameters by observables, such as pion mass and decay constant, and further to study the EoS of QCD at finite $\mu_B$. Thereafter, the EoS are employed to study the structures of compact stars. There are also some models fit quantities from LQCD at finite temperature~\cite{fan2017njl}. However, there is no reliable results from LQCD at finite baryon chemical potential.

In this paper, we will employ NJL model to fit the isospin density of recent LQCD data at finite isospin chemical potential, and then study the QCD phase structure in the $\mu_q-\mu_I$ plane. The EoS are calculated with the conditions of $\beta$ equilibrium and electric neutrality, and then take it as input to TOV equation to study the structure of neutron stars. The maximum mass and the radius of $1.4M_\odot$ of neutron stars are compared with the recent observations.

The paper is organized as follows. In Sec.~\ref{NJLmodel}, we give a basic introduction to the NJL model, which parameters are fitted by LQCD data. The phase structures in $\mu_q-\mu_I$ plane is studied in details. The EoS of QCD at finite $\mu_q$, finite $\mu_I$ and $T=0$ are discussed in Sec.~\ref{neutronStars}, thereafter the mass-radius relation for various central pressure are shown, which compared with recent observations. Finally, we give a brief summary in Sec.~\ref{sum}.
\section{NJL model and phase diagram in $\mu_B-\mu_I$ plane}\label{NJLmodel}
\subsection{Introduction to NJL model}
NJL model is an effective model of QCD in the low energy scale. The Lagrangian density of NJL is
\begin{eqnarray}
\mathscr{L}_\mathrm{NJL} &=& \mathscr{L}_0 + \mathscr{L}_{int},
\\
\mathscr{L}_0 &=& \bar\psi \left( i\slashed{\partial} -m \right)\psi,
\\
\mathscr{L}_{int} &=& G\left[ (\bar\psi\psi)^2 + (\bar\psi i\gamma_5\vec{\tau}\psi)^2 \right].
\end{eqnarray}
The second term is the interaction between quarks, which is non-renormalizable since dimension of the interaction are $[M^6]$. Therefore the cut-off is inevitable. In the two flavor case, $\psi=(u,d)^T$, and $\bar\psi=(\bar u, \bar d)$. The mass matrix for quarks is $m=\mathrm{diag}(m_u,m_d)$, we focus on the case of $m_u=m_d$ in this work. $G$ is the interaction strength. This form of interaction holds the chiral symmetry of QCD in the tree level, which enables us to study the chiral symmetry breaking and restoring phase transition.

In the mean field approximation, the interaction terms written as
\begin{eqnarray}
\bar\psi\psi &=& \sigma + \delta_\sigma,
\\
\delta_\sigma &=& \bar\psi\psi - \sigma,
\\
\bar\psi i\gamma_5\vec{\tau}\psi &=& \vec{\pi} + \vec\delta_\pi,
\\
\vec\delta_\pi &=& \bar\psi i\gamma_5\vec{\tau}\psi - \vec{\pi},
\end{eqnarray}
where $\sigma=\langle\bar\psi\psi\rangle$ and $\vec{\pi}=\langle\bar\psi i\gamma_5\vec{\tau}\psi\rangle$ are the average values of operators, $\bar\psi\psi$ and $\bar\psi i\gamma_5\vec{\tau}\psi$, in the ground state, respectively. The $\delta_\sigma$ and $\vec\delta_\pi$ are fluctuation of the two operators. The interaction can be written as
\begin{eqnarray}
\mathscr{L}_{int} &=& G\left[ (\bar\psi\psi)^2 + (\bar\psi i\gamma_5\vec{\tau}\psi)^2 \right]
\nonumber\\
&=& G\big[ \sigma^2 + 2\sigma \delta_\sigma + \delta_\sigma^2 + \vec{\pi}^2 +2\vec{\pi}\cdot \vec\delta_\pi + {\vec\delta_\pi}^2 
\big]
\nonumber\\
&\simeq& G\big[ \sigma^2 + 2\sigma \delta_\sigma + \vec{\pi}^2 +2\vec{\pi}\cdot \vec\delta_\pi
\big]
\nonumber\\
&=&  G\big[ \bar\psi\psi\sigma-\sigma^2
 +\bar\psi i\gamma_5\vec{\tau}\psi\cdot\vec{\pi} - {\vec{\pi}}^2
\big].
\end{eqnarray}
In this derivation, we have neglected the square of fluctuations.
After such approximation, namely mean field approximation, the interaction becomes bilinear of quark fields, which is solvable.

Concerning on the thermal and dense nuclear matter system, the thermodynamic potential per volume at finite temperature and chemical potential for both $u-$ and $d-$quarks~($\mu_u$ and $\mu_d$) is defined as
\begin{widetext}
\begin{eqnarray}
\Omega &=& -\frac{T}{V} \mathrm{Ln}\mathcal{Z}
\nonumber\\
&=& -\frac{T}{V} \mathrm{Ln} \mathrm{Tr} \exp\left( -\frac{1}{T}\int d^3x (\mathscr{H} - \bar\psi\mu\gamma^0\psi) \right)
\nonumber\\
&=&  -\frac{T}{V} \mathrm{Ln} \int\mathcal{D}\psi^\dagger\mathcal{D\psi} \exp\left[ \int_0^{1/T} d\tau \int d^3x \left( -\psi^\dagger\frac{\partial\psi}{\partial\tau} - \mathscr{H} + \bar\psi\mu\gamma^0\psi \right) \right]
\nonumber\\
&=&   -\frac{T}{V} \mathrm{Ln} \int\mathcal{D}\psi^\dagger\mathcal{D\psi} \exp\left[ \int_0^{1/T} d\tau \int d^3x \left(\mathscr{L}_\mathrm{NJL} + \bar\psi\mu\gamma^0\psi \right) \right],
\end{eqnarray}
\end{widetext}
where $\tau=i t$ is the imaginary time, $\mathcal{Z}$ is the grand canonical partition function, $\mathscr{H}$ is the Hamiltonian density, the trace take over all space, i.e., flavor, color, spin and momentum. The chemical potential matrix in the flavor space is
\begin{eqnarray}
\mu=\left(
\begin{array}{cc}
\mu_u   &0  \\
0       &\mu_d
\end{array}
\right)
=
\mu_q
\mathbf{I}
+\frac{1}{2}\mu_I
\tau_3
.
\end{eqnarray}
The relation between these chemical potentials are
\begin{eqnarray}
\mu_q &=& \frac{1}{2}\left( \mu_u + \mu_d \right),
\\
\mu_I &=& \mu_u - \mu_d.
\end{eqnarray}

In the mean field approximation,
\begin{eqnarray}
&&\mathscr{L}_\mathrm{NJL} + \bar\psi\mu\gamma^0\psi
\nonumber\\
&=& \bar\psi\left(i\slashed{\partial} - m + 2G\left(\sigma + i\gamma_5\vec{\tau}\cdot\vec{\pi} \right) +\mu\gamma^0\right)\psi - G(\sigma^2+\vec{\pi}^2)
\nonumber\\
&=& \bar\psi\left(i\slashed{\partial} - M + 2iG\gamma_5\vec{\tau}\cdot\vec{\pi} +\mu\gamma^0\right)\psi - G(\sigma^2+\vec\pi^2).
\end{eqnarray}
The $-G(\sigma^2+\pi^2)$ term is irrelevant to the fields in the thermodynamic potential, and hence can be factor out, the remain terms have the same form as the case of free Fermi-gas system,
\begin{eqnarray}
\Omega &=& \Omega_S + G(\sigma^2+\vec\pi^2),
\nonumber\\
\Omega_S &=& -T\sum_n \int\frac{d^3p}{(2\pi)^3} \mathrm{Tr}\mathrm{Ln}\left(\frac{1}{T} S^{-1}(i\omega_n,\vec{p})\right)
\nonumber\\
&=&-T\sum_n \int\frac{d^3p}{(2\pi)^3} \mathrm{Ln} \mathrm{Det}\left(\frac{1}{T} S^{-1}(i\omega_n,\vec{p})\right),
\end{eqnarray}
here
\begin{eqnarray}
&&S^{-1}(i\tilde\omega_n, \vec{p}) = 
\nonumber\\
&&
\left(
\begin{array}{cc}
\vec{\gamma}\cdot\vec{p} +i(\omega_n + i\mu_u)\gamma_0 - M  &G\pi_- i\gamma_5  \\
G\pi_+i\gamma_5   &\vec{\gamma}\cdot\vec{p} +i(\omega_n + i\mu_d)\gamma_0 - M
\end{array}
\right),
\nonumber\\
\end{eqnarray}
where $M=m-2G\sigma$ is the effective mass of dressed quark, $\omega_n=(2n+1)\pi T$ is the Matsubara frequency, $\pi_+$ and $\pi_-$ defined as
\begin{eqnarray}
\pi_+ &=& \langle\bar d i\gamma_5 u\rangle = \langle\bar d i\gamma_5(\tau_1+i\tau_2) u\rangle = \pi_1+i\pi_2,
\\
\pi_- &=& \langle\bar u i\gamma_5 d\rangle = \langle\bar d i\gamma_5(\tau_1-i\tau_2) u\rangle = \pi_1-i\pi_2.
\end{eqnarray}
These two quantities could be nonzero at finite isospin chemical potential, while $\pi_3$ keeps vanishing. After calculating the determinant of the inverse of quark propagator in all space, the thermodynamic potential finally is
\begin{eqnarray}
\Omega &=& G(\sigma^2+\vec\pi^2) -2N_c\int\frac{d^3k}{(2\pi)^3} \Big[ E_k^+ + E_k^- 
\nonumber\\
&&+ T\Big( \ln(1+e^{-(E_k^++\mu_q)/T}) + \ln(1+e^{-(E_k^+-\mu_q)/T}) 
\nonumber\\
&&+ \ln(1+e^{-(E_k^-+\mu_q)/T}) + \ln(1+e^{-(E_k^--\mu_q)/T}) \Big) \Big],
\nonumber\\
\end{eqnarray}
where $E_k^\pm$ defined as
\begin{eqnarray}
E_k^\pm &=& \sqrt{(E_k\pm\mu_I)^2+ 4G^2\vec\pi^2},
\\
E_k &=& \sqrt{\vec{k}^2+M^2}.
\end{eqnarray}
The physical quantities of the dressed quark propagator, namely $M$, $\pi_+$ and $\pi_-$, satisfy $\frac{\delta\Omega}{\delta M}=0$ and $\frac{\delta\Omega}{\delta \vec\pi}=0$, which make sure the thermodynamic potential located at minimum value. After some derivations, the equations of the dressed quark mass, $M$, and the pion condensate, $\langle\pi\rangle=\pi_++\pi_-$, are
\begin{eqnarray}
M &=& m - 2N_cG\int\frac{d^4k}{(2\pi)^4} i\mathrm{Tr}_D[S_{uu}(k) + S_{dd}(k)],
\\
\langle\pi\rangle &=& -2N_c G\int\frac{d^4k}{(2\pi)^4} \mathrm{Tr}_D[\left( S_{ud}(k) + S_{du}(k) \right)\gamma_5].
\end{eqnarray}
These two coupled equations can be solved by iteration.
\subsection{Phase diagram in $\mu_q-\mu_I$ plane}
\begin{figure}[!t]
\flushleft
\includegraphics[width=0.4\textwidth]{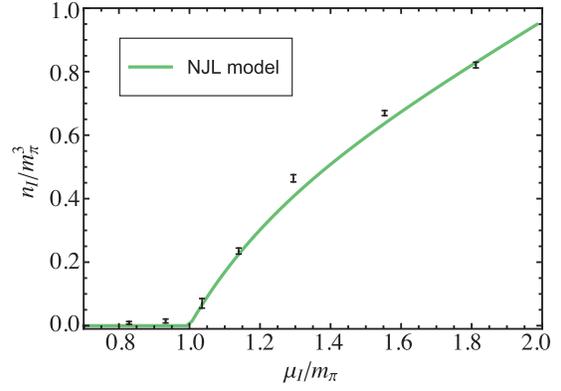}
\caption{The isospin density varying with $\mu_I$ at $\mu_q=0$ and compared with recent lattice QCD data~\cite{PhysRevD.98.094510}.}
\label{fig1}
\end{figure}
The quark number and isospin densities are crucial quantities for the QCD phase structures, they are defined as
\begin{eqnarray}
n_q &=& \langle u^\dagger u\rangle + \langle d^\dagger d\rangle
\nonumber\\
&=& -N_c \int\frac{d^4k}{(2\pi)^4} \mathrm{Tr}_D[i\left( S_{uu}(k) + S_{dd}(k) \right)\gamma^0],
\nonumber\\
n_I &=& \langle u^\dagger u\rangle - \langle d^\dagger d\rangle
\nonumber\\
&=& -N_c \int\frac{d^4k}{(2\pi)^4} \mathrm{Tr}_D[i\left( S_{uu}(k) - S_{dd}(k) \right)\gamma^0].
\end{eqnarray}
In this work, we employ parameters, $m=0.005$~GeV, $\Lambda=0.66$~GeV, $G=4.8~\mathrm{GeV}^{-2}$, which are fitted by pion mass, decay constant and the isospin density of recent lattice QCD study~\cite{PhysRevD.98.094510}. The isospin density varying with $\mu_I$ at $\mu_q$ is displayed in Fig.~\ref{fig1}. The $n_I$ keeps vanishing in the region of $\mu_I\le m_\pi=0.135~\mathrm{MeV}$, it continuously increase with $\mu_I$ which globally coincidence with lattice QCD data.

\begin{figure}[!b]
\flushleft
\includegraphics[width=0.46\textwidth]{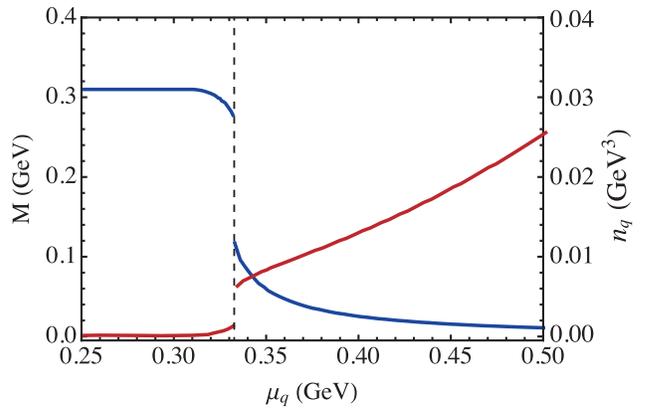}
\caption{The effective quark mass (dark blue line) with left axis and quark number density (dark red line) with right axis varying with $\mu_q$ at $\mu_I=0$.}
\label{fig2}
\end{figure}

At $\mu_I=0$, the $\langle\pi\rangle=0$ for any $\mu_q$, the effective mass of the dressed quark and quark number density varying with $\mu_q$ are displayed in Fig.~\ref{fig2}.
We can see that $M$ is a constant value in the region of $\mu_q<0.31~\mathrm{GeV}$, and thereafter $M$ smoothly decrease. At $\mu_q^c=0.333$~GeV, $M$ suffers a sudden change, then $M$ smoothly decrease in the region of $\mu_q>\mu_q^c$. The quark number density keeps zero for $\mu_q<0.31~\mathrm{GeV}$~(approximately one third of nucleon mass), and then $n_q$ smoothly increase, which implies baryons begin exist. The $n_q$ confronts the same sudden change at $\mu_q^c$, the $n_q$ increase smoothly after this critical point.

\begin{figure}[!t]
\centering
\includegraphics[width=0.4\textwidth]{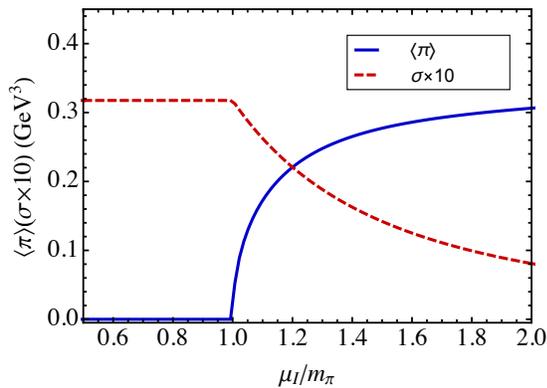}
\caption{The chiral and pion condensates varying with $\mu_I$.}
\label{fig3}
\end{figure}

At $\mu_q=0$ and finite $\mu_I$, the pion condensate $\langle\pi\rangle$ and chiral condensate varying with $\mu_I$ are shown in Fig.~\ref{fig3}, which are order parameters for pion superfluidity and chiral phase transitions respectively. 
Both of $\langle\pi\rangle$ and $\sigma$ are keep constants in the region of $\mu_I<m_\pi$, the $\langle\pi\rangle$ turns to nonzero at $\mu_I=\pi$ and suffers a rapid increase after this point which implies the pion superfluidity phase transition happens. The $\sigma$ begin decrease at $\mu_I=m_\pi$, and keeps nonzero upto $\mu_I=2m_\pi$.

\begin{figure}[!b]
\centering
\includegraphics[width=0.45\textwidth]{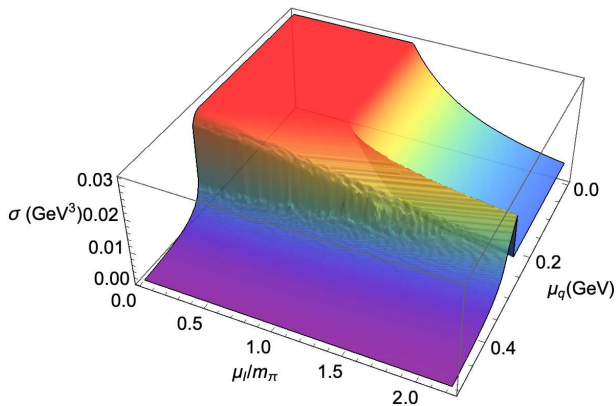}
\caption{Chiral condensate at $\mu_q-\mu_I$ plane.}
\label{fig4}
\end{figure}

\begin{figure}[!t]
\centering
\includegraphics[width=0.45\textwidth]{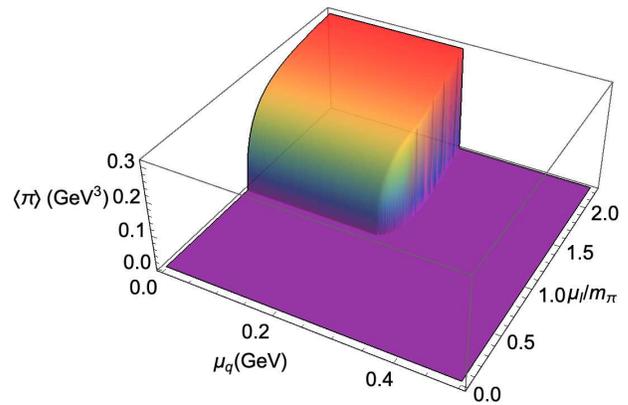}
\caption{Pion condensate at $\mu_q-\mu_I$ plane.}
\label{fig5}
\end{figure}

\begin{figure}[!h]
\centering
\includegraphics[width=0.45\textwidth]{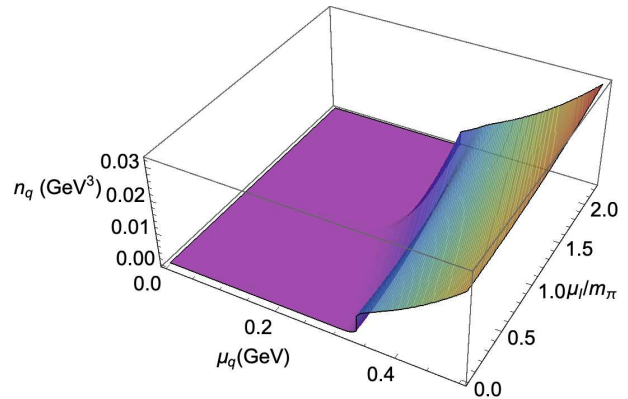}
\caption{Quark number density at $\mu_q-\mu_I$ plane.}
\label{fig6}
\end{figure}

The quantities of $\sigma$, $\langle\pi\rangle$ and $n_q$ are calculated in $\mu_q-\mu_I$ plane, they are shown in Fig.~\ref{fig4}, Fig.~\ref{fig5} and Fig.~\ref{fig6} respectively. In Fig.~\ref{fig4}, we can see that the $\sigma$ is a constant in the low $\mu_q$ and $\mu_I$ region. When $\mu_I>m_\pi$, the $\sigma$ suffers two discontinuity during $\mu_q$ increase.

In Fig.~\ref{fig5}, the $\langle\pi\rangle=0$ in $\mu_I<m_\pi$ for any $\mu_q$, it continuously turns to nonzero at $\sim m_\pi$ for $\mu_q<0.27$~GeV, which implies the second order phase transition happens. For $\mu_q>0.27$~GeV, the $\langle\pi\rangle$ jumps from zero to a nonzero value, which indicates the first order phase transition happens. The junction of them located at $(\mu_q,\mu_I)=(0.270,0.158)$~GeV, which is a tri-critical point~(TCP).

The quark number density $n_q$ is an important quantity to understand the structure of compact stars. In Fig.~\ref{fig6}, we can see that the $n_q\neq 0$ appears at $\mu_q=310$~MeV$\sim m_N/3$ for $\mu_I=0$, which has shown in Fig.~\ref{fig2}. During $\mu_I$ increase, the critical $\mu_q^c(\mu_I)$ for the quark number density decrease.

\section{Neutron stars}\label{neutronStars}
\subsection{Equation of state}
The compact stars are almost electric neutral celestial bodies because of $\beta$ equilibrium, which give constrains for the densities of electron, $u-$ and $d-$quarks. The constrains are
\begin{eqnarray}
\mu_d&=&\mu_u+\mu_e,
\\
\frac{2}{3}n_u &=& \frac{1}{3}n_d + n_e,
\end{eqnarray} 
\begin{figure}[!t]
\centering
\includegraphics[width=0.4\textwidth]{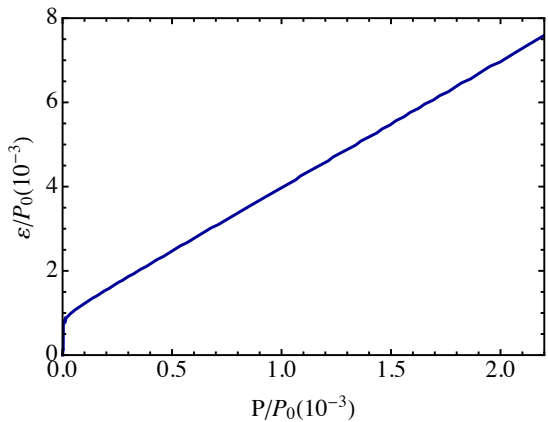}
\caption{The EoS of the dense QCD system with $\beta$ equilibrium and electric neutrality conditions.}
\label{fig7}
\end{figure}
where $\mu_u$, $\mu_d$ and $\mu_e$ are chemical potentials for $u$ quark, $d$ quark and electron, $n_u$, $n_d$ and $n_e$ are their particle densities. Particle number densities are depend on their chemical potentials, therefore there is only one independent variable in these chemical potentials. One can choose $\mu_u$ as the independent variable without loss of generality. The relation between pressure and thermodynamic potential is
\begin{eqnarray}
P(\mu_q,\mu_I) = -\Omega(\mu_q,\mu_I).
\end{eqnarray}
The energy density $\varepsilon$ is
\begin{eqnarray}
\varepsilon = -P + \sum_{i} \mu_i n_i,
\end{eqnarray}
where summation take over all kinks of particles, namely electron and $u-$, $d-$quarks. The relation between pressure and energy density is displayed in Fig.~\ref{fig7}, where $P_0$ is defined in Eq.~(\ref{P0}).
\subsection{Mass-radius relation of neutron stars}
The structure of neutron stars can be obtained by solving coupled equations
\begin{eqnarray}
\frac{dP}{dr} &=& - \frac{G(\varepsilon + P)(M + \frac{4\pi r^3 P}{c^2})}{r(rc^2-2GM)},    \label{TOV}
\\
\frac{dM}{dr} &=& 4\pi r^2 \frac{\varepsilon}{c^2}. \label{mass}
\end{eqnarray}
We define dimensionless quantities, $\overline{M},\bar r, \bar P, \bar\varepsilon$, as follows
\begin{eqnarray}
M&=&\overline{M}M_\odot,
\\
r&=&\bar{r}R_0,\qquad R_0= \frac{GM_\odot}{c^2}=1.477~\mathrm{km},    \label{R0}
\\
P&=&\bar{P}P_0, \qquad \varepsilon =\bar\varepsilon P_0,
\\
P_0&=&\frac{M_\odot c^2}{R_0^3}=5.547\times 10^{37}~\mathrm{Pa}
\nonumber\\
&&\hspace*{10.5mm}=346.255~\mathrm{GeV}\cdot\mathrm{fm}^{-3}.   \label{P0}
\end{eqnarray}
The Eqs.~(\ref{TOV}) and (\ref{mass}) become
\begin{eqnarray}
\frac{d\bar{P}}{d\bar{r}} &=& - \frac{\left( \bar\varepsilon + \bar P \right)(\overline{M} + \bar{r}^3\bar{P})}{\bar r(\bar r -2\overline{M})},
\\
\frac{d\overline{M}}{d\bar r} &=& 4\pi \bar r^2 \bar\varepsilon.
\end{eqnarray}
\begin{figure}[!h]
\centering
\includegraphics[width=0.4\textwidth]{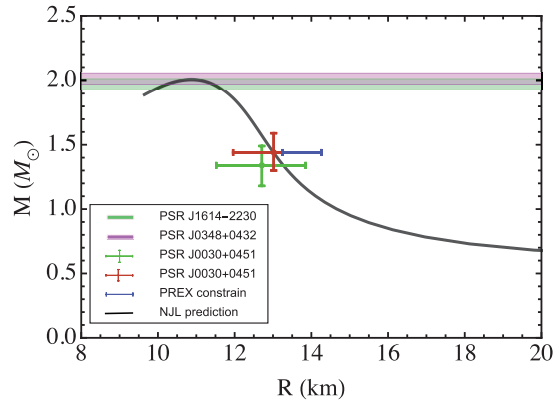}
\caption{The mass and radius relation of neutron stars.}
\label{fig8}
\end{figure}
Take the EoS in Fig.~\ref{fig7} as an input, the above coupled equations can be solved in numerical with boundary conditions, $P(r=0)=P_c$ as a given choice of central pressure and $M(r=0)=0$. At the radius of star, $P(r=R)=0$ and $M(r=R)$ is the total mass of the star. Mass-radius relation is plotted in Fig.~\ref{fig8}, some recent observed neutron stars are dispalyed as comparison~\cite{Demorest2010,Arzoumanian_2018,Antoniadis1233232,riley2019nicer,miller2019psr,PhysRevLett.126.172503}. In Fig.~\ref{fig8}, we can see that the maximum mass is roughly $2M_\odot$. The observation of the binary millisecond pulsar J1614-2230 report the mass to be $(1.97\pm0.04)M_\odot$~\cite{Arzoumanian_2018}, which is shown in Fig.~\ref{fig8}. Another observation of the pulsar J0348+0432 report the mass of $(2.01\pm 0.04)M_\odot$~\cite{Antoniadis1233232}. They are only two neutron stars with precisely determined mass of $\sim 2M_\odot$, which are very close to our prediction of the upper limit of neutron stars. In addition, the recent observations of $1.4M_\odot$ neutron stars are compared with our predictions in Fig.~\ref{fig8}. The inferred mass and equatorial radius of pulsar PSR J0030+0451 are $1.34_{-0.16}^{+0.15}M_\odot$ and $12.71_{-1.19}^{+1.14}~\mathrm{km}$ respectively~\cite{riley2019nicer}, which is close to the our predicted curve. And another group give result as $M=1.44_{-0.14}^{+0.15}M_\odot$ and $R=13.02_{-1.06}^{+1.24}~\mathrm{km}$~\cite{miller2019psr}, the central values are exactly coincident with what we predicted. It is worth noting that the PREX collaboration report their recent radius constrains of neutron stars with mass of $1.4M_\odot$, that is $13.25~\mathrm{km}\lesssim R^{1.4M_\odot}\lesssim 14.26~\mathrm{km}$~\cite{PhysRevLett.126.172503}, which is much larger than many previous results~\cite{capano2020stringent,baubock2015rotational} and very close to our prediction.

\section{Summary}\label{sum}
The parameters of NJL model are fitted by pion mass and decay constant together with isospin density from lattice QCD. The QCD phase structures in $\mu_q-\mu_I$ plane are studied in detail. At $\mu_I=0$, the nonzero quark number density appears at $\mu_q=0.31~\mathrm{GeV}\sim m_N/3$, which is consistent with our usual understanding of nuclear matter. With the increase of $\mu_q$, the nuclear matter suffered a chiral phase transition at $\mu_q=0.333~\mathrm{GeV}$, and there is a sudden increase of the quark number density. At $\mu_q=0$ and $\mu_I\neq 0$, the QCD systems undergoes a pion-superfluidity phase transition at $\mu_I=m_\pi$. The quark condensate continuously changes and keeps nonzero upto $\mu_I=2m_\pi$, which implies chiral symmetry is not restored. Thereafter, the $\sigma$, $\langle\pi\rangle$ and $n_q$ are plotted at $\mu_q-\mu_I$ plane.

Take the $\beta$ equilibrium and electric charge neutrality of neutron star into account, the chemical potentials, $\mu_u$, $\mu_d$ and $\mu_e$, are not independent. There is only one independent variable, we choose $\mu_u$ as the independent variable without loss of generality. The equation of states are displayed with the help of the phase structures of dense QCD matter. Take the EoS as input, the TOV equation is solved numerically with specific central pressure. The results show that the maximum mass of neutron stars is around $2M_\odot$, which is close to the maximum observed neutron stars, such as pulsar J1614-2230 and pulsar J0348+0432. The recent observations of PSR J0030+0451 are consistent with our mass-radius relation. Especially, the PREX collaboration report their recent radius constrains of $1.4M_\odot$ neutron stars, which is much larger than many previous studies, $\sim 10~\mathrm{km}$, while it is close to our prediction.

\acknowledgments
This work is supported in part by the National Natural Science Foundation of China (under Grant No. 11905107), the National Natural Science Foundation of Jiangsu Province of China (under Grant No. BK20190721), Natural Science Foundation of the Jiangsu Higher Education Institutions of China (under Grant No. 19KJB140016), Nanjing University of Posts and Telecommunications Science Foundation (under grant No. NY129032), Innovation Program of Jiangsu Province.
\bibliography{references}
\end{document}